\documentclass[11pt,twoside]{article}
\usepackage[]{graphicx}
\usepackage{amsmath}
\usepackage{amssymb}
\usepackage{cite}
\usepackage[cp1250]{inputenc}

\begin{document}
\newcommand{\pst}{\hspace*{1.5em}}


\newcommand{\be}{\begin{equation}}
\newcommand{\ee}{\end{equation}}
\newcommand{\bm}{\boldmath}
\newcommand{\ds}{\displaystyle}
\newcommand{\bea}{\begin{eqnarray}}
\newcommand{\eea}{\end{eqnarray}}
\newcommand{\ba}{\begin{array}}
\newcommand{\ea}{\end{array}}
\newcommand{\arcsinh}{\mathop{\rm arcsinh}\nolimits}
\newcommand{\arctanh}{\mathop{\rm arctanh}\nolimits}
\newcommand{\bc}{\begin{center}}
\newcommand{\ec}{\end{center}}

\thispagestyle{plain}

\label{sh}


\begin{center} {\Large \bf
\begin{tabular}{c}
ANGULAR UNCERTAINTY OF MOMENTUM CORRELATIONS
\\[-1mm]
IN PARAMETRIC FLUORESCENCE
\end{tabular}
 } \end{center}

\bigskip

\bigskip

\begin{center} {\bf
Martin Hamar, Jan Pe\v{r}ina Jr., Václav Michálek and Ond\v{r}ej Haderka$^*$
}\end{center}

\medskip

\begin{center}
{\it
Joint Laboratory of Optics of Palack\'{y} University
and Institute of Physics of Academy of Sciences of the Czech
Republic, 17. listopadu 50a, 772 07 Olomouc, Czech Republic
}
\smallskip

$^*$Corresponding author e-mail:~~~ondrej.haderka~@~upol.cz\\
\end{center}

\begin{abstract}\noindent
Uncertainty in the determination of emission angles of signal and
idler photons is investigated. The role of divergence of the pump
beam, width of the pump spectrum, and size of the crystal is
elucidated. Experimental data obtained by an iCCD camera are in a
good agreement with a numerical model that provides angular and
spectral characteristics of the signal and idler fields.
\end{abstract}

\medskip

\noindent{\bf Keywords:} photon pairs, uncertainty, correlation,
entanglement

\section{Introduction}
\pst Light emitted from parametric fluorescence in a nonlinear
crystal is composed of photon pairs. During the emission
frequencies and emission directions of two photons comprising a
pair are determined by the laws of energy and momentum
conservation, as has been recognized already in year 1968
\cite{Giallorenziho1968}. For this reason, there occurs a strong
correlation (entanglement) between properties of the signal and
idler photons. In an ideal case of infinitely long and wide
nonlinear crystal and monochromatic plane-wave pumping, a
plane-wave signal photon at frequency $ \omega_s $ has just one
plane-wave idler photon at frequency $ \omega_i $. Emission angles
of two photons are given by momentum conservation that forms
phase-matching conditions. Possible signal (and similarly idler)
emission directions lie on a cone which axis coincides with the
pump-beam direction of propagation. In real experimental
conditions, blurring of emission directions occurs because of
crystals of finite dimensions \cite{Hong1985,Wang1991}, pump-beam
divergence \cite{Grayson1994,Steuernagel1998} as well as pulsed
pumping \cite{Keller1997,PerinaJr1999}. Accepting the
approximation based on a multidimensional gaussian two-photon
wave-function an analytically-tractable theoretical model has been
developed \cite{Joobeur1994,Joobeur1996}. Recently, theoretical
models have been generalized to photonic \cite{PerinaJr2006} and
wave-guiding \cite{PerinaJr2008} structures.

Here we extend the previous investigations of photon-pair
properties by studying correlation areas of the signal and idler
photons. Experimental results are compared with a theoretical
model that considers Gaussian temporal spectrum and elliptical
pump-beam profile.

Model of parametric fluorescence is briefly described in Sec.~2.
Details of the experimental setup are revealed in Sec.~3. The
obtained experimental results are discussed in Sec.~4. Sec.~5
provides conclusions.

\section{Model of parametric fluorescence}
\pst The process of parametric fluorescence is described by the
following interaction Hamiltonian $ \hat{H}_{\rm int} $
\cite{Hong1985,Shih2003}:
\begin{eqnarray}      
 \hat{H}_{\rm int}(t) &=& \varepsilon_0 \int\limits_V d{\bf r}
\, \chi^{(2)} : {\bf E}_{p}^{(+)}({\bf r},t) \hat{\bf E}_{s}^{(-)} ({\bf r},t)
 \hat{\bf E}_{i}^{(-)}({\bf r},t) + {\rm H.c.} ,
\label{1}
\end{eqnarray}
where $ {\bf E}_{p}^{(+)} $ is the positive-frequency part of the
pump-field electric-field amplitude, whereas $ {\bf E}_{s}^{(-)} $
($ {\bf E}_{i}^{(-)} $) stands for the negative-frequency part of
the signal- (idler-) field electric-field amplitude operator.
Symbol $ {\bf \chi^{(2)}} $ denotes the second-order
susceptibility tensor and $ : $ is shorthand for tensor reduction
with respect to its 3 indices. Susceptibility of vacuum is denoted
as $ \varepsilon_0 $, interaction volume as $ V $ and $ {\rm H.c.}
$ substitutes Hermitian-conjugated terms.

Parametric fluorescence of type-I in an LiIO$ {}_3 $ crystal
oriented such that its optical axis was perpendicular to the $ z $
axis of fields' propagation was investigated. The pump field was
polarized vertically (it propagated as an extraordinary wave)
whereas the signal and idler fields propagated horizontally
polarized (as ordinary waves). Under these conditions, a scalar
theory is sufficient for the description. Because of low signal-
and idler-field intensities, the Schr\"{o}dinger equation can be
solved only to the first order \cite{PerinaJr1999} and this
solution then provides the fourth-order correlation function of
the signal- and idler-field electric-field amplitudes $
G^{(2)}_{s,i} $ in the form:
\begin{equation}     
 G_{s,i}^{(2)}(\mathbf{k}_{s0}, \mathbf{k}_{i0}) =
 \int_{\Delta {\bf k}_s}  d^3 \mathbf{k}_{s} \int_{\Delta {\bf
 k}_i} d^3 \mathbf{k}_{i} \left| d(\mathbf{k}_{s})
 d(\mathbf{k}_{i})\right|^2 \left|  S ( \mathbf{k}_{s},
 \mathbf{k}_{i}) \right|^2,
\label{2}
\end{equation}
where $d(\mathbf{k}_{s}) $ [$ d(\mathbf{k}_{i}) $] is amplitude
transmissivity of a detector for the signal [idler] field plane
wave with wave vector $ {\bf k}_s $ [$ {\bf k}_i $]. The
correlation function $ G^{(2)}_{s,i} $ gives the number of photon
pairs generated with a signal-photon wave vector $ {\bf k}_s $ in
the area defined by $ \Delta {\bf k}_s $ around $ {\bf k}_{s0} $
and the idler-twin wave vector $ {\bf k}_i $ in the area around $
{\bf k}_{i0} $ described by $ \Delta {\bf k}_i $. Correlation
function $ S $ introduced in Eq.~(\ref{2}) can be expressed in the
form
\begin{equation}   
 S( \mathbf{k}_{s}, \mathbf{k}_{i}) = C_n \int d^3 \mathbf{k}_{p }
 E_{p}^{( + )}(\mathbf{k}_{p}) \delta (\omega_s  + \omega_i  - \omega_p)
 \int\limits_V d^3 \mathbf{r} \exp\left[-i( \mathbf{k}_{p}  -
 \mathbf{k}_{s}  - \mathbf{k}_{i }) \mathbf{r} \right]
\label{3}
\end{equation}
that reflects the role of the pump beam in the determination of
the signal- and idler-field properties. Physically, it gives the
probability amplitude of simultaneous emission of a signal photon
with wave vector $\mathbf{k}_{s}$ together with an idler photon
having wave vector $\mathbf{k}_{i}$. Symbol $ C_n $ in
Eq.~(\ref{3}) stands for a normalization constant. Integrals
occurring in the expressions in Eqs.~(\ref{2}) and (\ref{3}) are
complex in general and require numerical approach. Program that
determines the correlation function $ G^{(2)}_{s,i} $ for any
type-I parametric fluorescence has been developed to obtain
results needed for comparison with the obtained experimental data.

\section{Experimental setup}
\begin{figure}[tb]  
\bc
\includegraphics[scale=0.7]{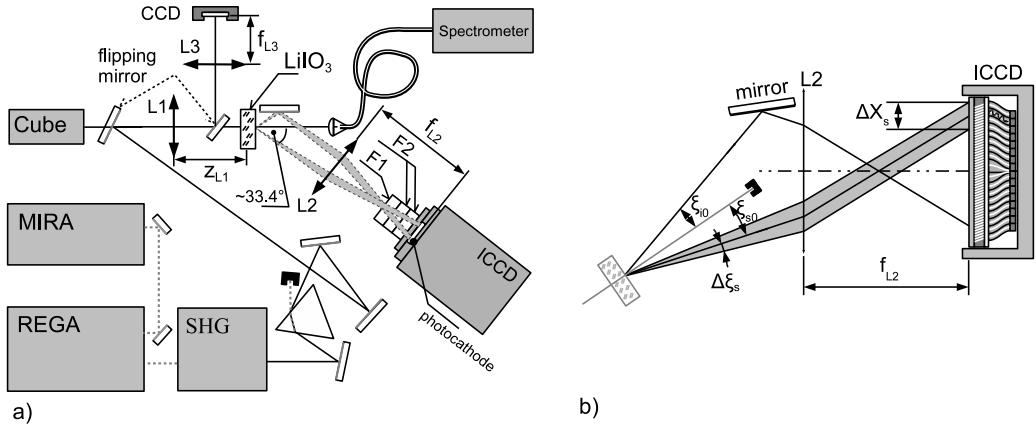}
\ec \caption{Experimental setup used for the measurement of
 angular uncertainties: (a) Whole setup using both
 pumping options and diagnostics of the pump beam (see the text for
 detailed description); (b) Part of the setup mapping
 propagation angles behind the nonlinear crystal to the positions
 on the photocathode.}
\label{fig1}
\end{figure}

\pst In our experiment, a negative uniaxial crystal made of
LiIO$_3$ (manufactured by EKSMA Optics) was cut for non-critical
phase matching (optical axis perpendicular to the pumping beam).
Two crystal lengths were used $L_{cr}$=2~mm or 5~mm both for cw
and pulsed pumping. In cw case, a semiconductor laser Cube 405
(Coherent) delivered 31.6~mW at 405~nm with spectral bandwidth
$\Delta \lambda=$0.59~nm (all widths are given as half-widths at
$1/e$ of the maximum). For pulsed pumping the second-harmonic
field of an amplified femtosecond Ti:sapphire system (Mira+RegA,
Coherent) lasing at 800~nm was used. The pulses were $\sim$250~fs
long at the fundamental wavelength. At a repetition rate of 11~kHz
the mean SHG power was 2.5~mW at the crystal input. Spectral
bandwidth could be adjusted between 1.7 and 2.6~nm using fine
tuning of the SHG process. The SHG beam was separated from the
fundamental beam using a dispersion prism (see Fig.~\ref{fig1}).

Divergence of both pumping options was controlled using
interchangeable converging lens L1 or beam expander (BE2X,
Thorlabs). Focal lengths $f_{L1}$ in the range from 30 to 75~cm
were used. Distance $z_{L1}$ between the lens L1 and the nonlinear
crystal was set such that the beam waist was placed behind the
crystal, $z_{L1}<f_{L1}$. Spatial spectrum of the pump beam was
measured by a CCD camera (Li085M, Lumenera) placed at a focal
plane of a converging lens L3. From the spatial spectra, vertical
and horizontal projections of the beam waist $W_{0l},\,l=x,y$,
were determined. Temporal spectrum of the pump beam was measured
by a fiber spectrometer (HR4000CG-UV-NIR, Ocean Optics) placed
behind the crystal.

\begin{figure}[tb]   
\bc
\includegraphics[scale=0.7]{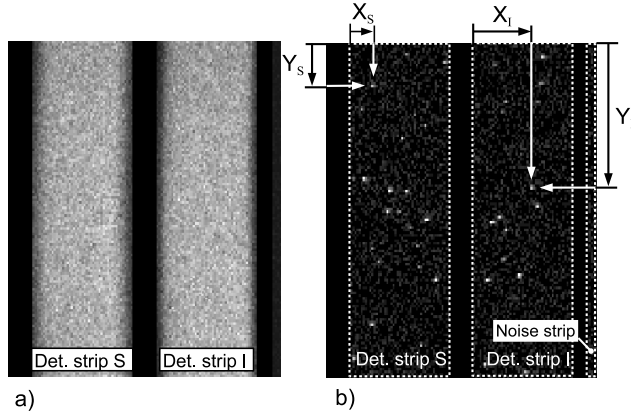}
\ec \caption{(a) Cumulative registration of photons produced by
 20,000 consecutive pump pulses in one camera frame. Signal and
 idler strips image small sections of the cone layer (note slight
 curvature of the strips, the curvature is oriented in the same
 sense for both strips because the idler strip is reflected from a
 mirror). (b) A typical frame containing photon detection events
 caused by one pump pulse.}
\label{fig2}
\end{figure}

In our experiment we focused on degenerate photon pairs $\lambda
_{s0}=\lambda _{i0} = 800$~nm that were emitted on a cone layer
with apex angle 33.4~deg behind the crystal. One section of the
cone layer was captured directly by the detector, the opposite
section was directed to the detector using a high-reflectivity
mirror (see Fig.~\ref{fig2}a). The detector was composed of an
iCCD camera with image intensifier (PI-MAX:512-HQ, Princeton
Instruments) preceded by a converging lens L2, one
narrow-bandwidth and two high-pass edge filters. The purpose of
lens L2 is to map the photon propagation direction angles to
positions at the photocathode (see Fig.~\ref{fig2}b). Several
lenses L2 with focal lengths $f_{L2}$= 12.5, 15, or 25~cm were
used in different variants of the experiment. Lens L2 was placed
such that the photocathode lied in its focal plane. Filter 11~nm
wide was centered at 800~nm. Edge filters (Andover, ANDV7862)
blocked wavelengths below 666~nm while their transmittance at
800~nm reached 98\%.

Photosensitive area of the detector in the form of a rectangular
shape 12.36~mm wide was divided into $512\times 512$ pixels.
Resolution of the camera was limited to 38~$\mu$m (FWHM) mainly
due to imperfect contrast transfer in the image intensifier. To
speed-up the data collection we decreased the resolution even
further by grouping $4\times 4$ or $8\times 8$ pixels into one
super-pixel in the hardware of the camera. Typically we captured
several tens of camera frames per second. Quantum efficiency of
the detector was estimated to 7\% including all the components
between the nonlinear crystal and photocathode.

To measure the signal-idler correlation function given in
Eq.~(\ref{2}) we defined three regions-of-interest at the iCCD
photocathode (see Fig.~\ref{fig2}). Two of them capture signal and
idler photons and their widths are given by bandwidth filter and
lens L2 focal length. The third strip serves for monitoring the
noise level. In a pulsed regime a 10~ns long gate of the camera
was used synchronously with laser pulses. Considering cw pumping
the camera was triggered internally with a gate lasting 2~$\mu$s.
Pump-field intensity was set so that the average number of photon
detection events per signal/idler strip was much lower than the
number of super-pixels in each strip ($\sim 5000$). This made the
probability of detecting two photons in a single super-pixel
negligible.

Not all detection events were due to signal or idler photons. We
made a detailed analysis of the noise and found out that 1.82\% of
detections came from noise the majority of which (9/10) were red
photons coming from fluorescence in the crystal. The rest were
scattered pumping photons and only 1.6\% of the noise was caused
by dark counts.

Evaluation of the detection frames allowed us to compute the
experimental correlation function in the horizontal ($X$) and
vertical ($Y$) coordinates at the camera (see Fig.~\ref{fig2}b) as
follows:
\begin{equation}  
 g\left( {X_S ,X_I } \right) =
  \sum\limits_{p = 1}^N {} \sum\limits_{m = 1}^{M_p } {} \sum\limits_{l = 1}^{L_p }
  {\frac{1}{{M_p L_p }}}\, \delta \left( {X_S ^{pm}  - X_S }
  \right)\delta \left( {X_{I} ^{pl}  - X_I } \right),
\label{4}
\end{equation}
where $p$ is an index of the frame ($N$ gives the number of
frames) and $m$ ($l$) counts signal (idler) detection events [up
to $M_p$ ($L_p$) in the $p$-th frame]. Symbol $X_s^{pl}$
[$X_i^{pl}$] denotes horizontal position of the $l$-th detection
in the signal [idler] strip in the $p$-th frame. Correlations in
the vertical direction are determined similarly. In this way, all
possible combinations of pairwise detection events are taken into
account.

\section{Results }
\begin{figure}   
\bc
\includegraphics[scale=0.7]{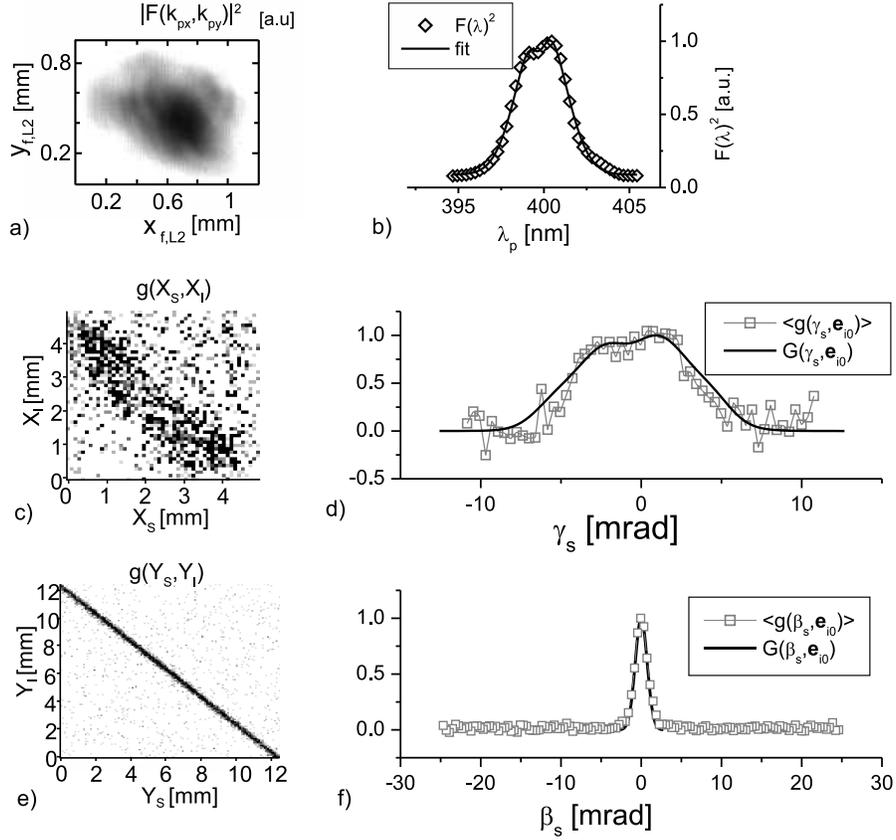}
\ec
\caption{Results of a typical measurement sequence with pulsed
 pumping containing 327,600 frames, $L_{cr}=5$~mm. (a) Spatial
 spectrum of the pump beam as measured in lens L3 focal plane. (b)
 Intensity spectrum of the pump beam; diamonds are experimental
 values from spectrometer, solid line is a multi-peak Gaussian fit.
 (c,e) Experimental correlation functions $g(X_S,X_I)$ and
 $g(Y_S,Y_I)$ evaluated according to Eq.~(\ref{4}). (d,f)
 Cross-sections of functions $g(X_S,X_I)$ and $g(Y_S,Y_I)$ along
 the signal axis averaged over all registered idler values (open
 rectangles) and the corresponding theoretical curves $G$ given by
 Eq.~(\ref{2}) obtained from the numerical model (solid lines).
 These plots are cast in angular units $\gamma_s \approx X_s/f_{L2}
 $, $ \beta_s \approx Y_s/f_{L2}$.}
\label{fig3}
\end{figure}

\pst A typical result for pulsed pumping based on a measurement
sequence lasting three hours is given in Fig.~\ref{fig3}.
Spatial spectrum of the pump beam as measured in the focal plane
of lens L3 is shown in Fig.~\ref{fig3}a and the corresponding
intensity spectral profile in Fig.~\ref{fig3}b. Plots of the
correlation functions $g(X_S ,X_I )$ and $g(Y_S ,Y_I )$ evaluated
from 327,000 registered frames according to Eq.~(\ref{4}) are
given in Figs.~\ref{fig3}c and \ref{fig3}e. Pairwise correlated
character of the fields leads to the diagonal patterns going from
upper-left to lower-right corners of the plots. The diagonals have
finite widths originating in spreading of signal propagation
directions related to one fixed idler direction. We denote
half-widths of the diagonals as $\Delta X_{s}/2$ and $\Delta
Y_{s}/2$ or, in angular quantities, as $\Delta \gamma_{s}/2
\approx \Delta X_{s}/(2 f_{L2})$ and $\Delta \beta_{s}/2 \approx
\Delta Y_{s}/(2 f_{L2})$. Since these widths apparently do not
change significantly in the $X $ and $ Y$ ranges used for plotting
Figs.~\ref{fig3}c and \ref{fig3}e, we can increase the precision
in determining $\Delta \gamma_{s}$ and $\Delta \beta_{s}$ by
averaging over the range of measured idler coordinates. The
averaged cross-sections along signal coordinates $\gamma_s $ and $
\beta_s$ are indicated by angle brackets $\langle \rangle$ and are
plotted in Figs.~\ref{fig3}d and \ref{fig3}f (open rectangles).
Here, solid lines show the corresponding curves giving
$G(\mathbf{k}_{s}, \mathbf{k}_{i0})$ that are obtained from the
numerical model using parameters of the pump beam as were derived
from the measurements reported in Figs.~\ref{fig3}a and
\ref{fig3}b. Excellent agreement of the model with experimental
data is evident even though the pump-field spatial shape and its
spectrum are nontrivial.
\begin{figure}   
\bc
\includegraphics[scale=0.8]{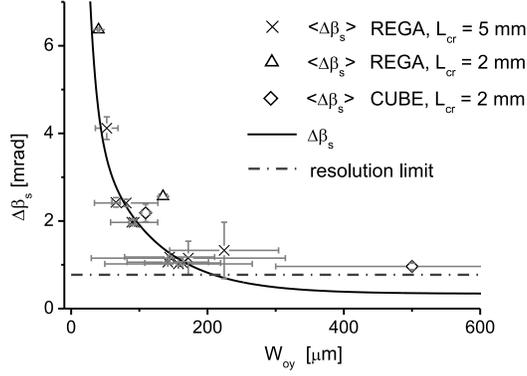}
\ec \caption{Dependence of vertical angular uncertainty
 $\langle\Delta \beta_{s}\rangle$ on the bandwidth of vertical
 pump-beam waist $W_{y0}$ for pulsed (crosses and triangles) and cw
 (diamonds) pumping. Two different crystal lengths were used: 5~mm
 (crosses) or 2~mm (triangles and diamonds). Solid line is provided
 by the numerical model. Dash-dot horizontal line indicates the
 resolution limit given by camera super-pixel size.}
\label{fig4}
\end{figure}

We have performed systematic investigation of the dependencies of
angular uncertainty $\langle\Delta \beta_{s}\rangle$ on various
parameters of the pump beam and crystal length. The role of
vertical dimension of the pump-beam waist $W_{y0}$ in the behavior
of the uncertainty $\langle\Delta \beta_{s}\rangle$ is revealed in
Fig.~\ref{fig4} where values for both pumping options and two
different crystal lengths are plotted on the top of the solid line
given by the numerical model. We note that in this case and
according to the model values of the uncertainty $\langle\Delta
\beta_{s}\rangle$ do not change with pump spectral width and
horizontal beam-waist size $W_{x0}$. We can see in Fig.~\ref{fig4}
that experimental points follow the theoretical curve except for
the area of narrow diagonals where the experimental resolution was
limited by the size of super-pixels.

\section{Conclusions}
\pst Thanks to the ability of the intensified CCD camera to
register single-photon detection events and positions of their
occurrence, we were able to measure angular uncertainties of the
far-field signal-idler correlations in the process of parametric
fluorescence in an LiIO$_3$ nonlinear crystal. In particular, we
have measured the dependence of vertical angular uncertainty on
the vertical pump-beam waist size using both cw and pulsed pumping
and two different crystal lengths. We have reached a very good
agreement with the numerical model of parametric fluorescence that
simulates the process for any polychromatic elliptical pump beam.

\section*{Acknowledgments}
\pst
This research was supported by the projects COST 09026, 1M06002 and
MSM6198959213 of the Ministry of Education of the Czech Republic.


\begin{thebibliography}{99}

\bibitem{Giallorenziho1968} T. G. Giallorenziho, C. L. Tang, Phys. Rev. \textbf{166}, 225 (1968).
\bibitem{Hong1985} C. K. Hong, L. Mandel, Phys. Rev. A \textbf{31}, 2409 (1985).
\bibitem{Wang1991} L. J. Wang, X. Y. Zou, L. Mandel, Phys. Rev. A \textbf{44}, 4614 (1991).
\bibitem{Grayson1994} T. P. Grayson, G. A. Barbosa, Phys. Rev. A. \textbf{49}, 2948
(1994).
\bibitem{Steuernagel1998} O. Steuernagel, Rabitz H., Opt. Commun. \textbf{154}, 285 (1998).
\bibitem{Keller1997} T. E. Keller, M. H. Rubin, Phys. Rev. A \textbf{56}, 1534 (1997).
\bibitem{PerinaJr1999} J. Peřina, Jr., A. V. Sergienko, B. M. Jost, B. E. A. Saleh,
M. C. Teich, Phys. Rev. A \textbf{59}, 2359 (1999).
\bibitem{Joobeur1994} A. Joobeur, B. E. A. Saleh, M. C. Teich, Phys. Rev. A \textbf{50}, 3349
(1994).
\bibitem{Joobeur1996} A. Joobeur, B. E. A. Saleh, T. S. Larchuk, M. C. Teich, Phys. Rev. A \textbf{53}, 4360
(1996).
\bibitem{Shih2003} Y. Shih, Rep. Prog. Phys. \textbf{66}, 1009 (2003).
\bibitem{PerinaJr2006} J. Peřina, M. Centini, C. Sibilia, M. Bertolotti, M. Scalora,
Phys. Rev A \textbf{73}, 033823 (2006).
\bibitem{PerinaJr2008} J. Peřina Jr., Phys. Rev A \textbf{77}, 013803 (2008).

\end{thebibliography}
\end{document}